\documentclass[aps,prx,twocolumn,superscriptaddress]{revtex4-1}

\usepackage[utf8]{inputenc}
\usepackage[T1]{fontenc}
\usepackage{amsmath}
\usepackage{amssymb}
\usepackage{amsfonts}
\usepackage{bm}
\usepackage{color}
\usepackage{datetime}
\usepackage{graphicx}
\usepackage[colorlinks,citecolor=blue,urlcolor=blue,linkcolor=blue]{hyperref}
\usepackage{physics}
\usepackage{mathtools}
\usepackage{enumerate}

\DeclareMathSymbol{*}{\mathbin}{symbols}{"01}

\newcommand{\expv}[1]{\expval*{#1}}

\newcommand{\id}{\mathrm{id}}

\begin{document}

\title{Data-driven approach to mixed-state multipartite entanglement characterisation}

\author{Eric Brunner}
\altaffiliation[Current address: ]{Quantinuum, Partnership House, Carlisle Place, London SW1P 1BX, United Kingdom}
\email{eric.brunner@quantinuum.com}
\affiliation{\freiburg}
\affiliation{\eucor}
\author{Aaron Xie}
\affiliation{\freiburg}
\author{Gabriel Dufour}
\affiliation{\freiburg}
\affiliation{\eucor}
\author{Andreas Buchleitner}
\affiliation{\freiburg}
\affiliation{\eucor}

\newcommand{\freiburg}{Physikalisches Institut, Albert-Ludwigs-Universit\"{a}t Freiburg, Hermann-Herder-Stra{\ss}e 3, 79104, Freiburg, Germany}
\newcommand{\eucor}{EUCOR Centre for Quantum Science and Quantum Computing, Albert-Ludwigs-Universit\"{a}t Freiburg, Hermann-Herder-Stra{\ss}e 3, 79104, Freiburg, Germany}

\date{\today}


\begin{abstract}

We develop a statistical framework, based on a manifold learning embedding, to extract relevant features of multipartite entanglement structures of mixed quantum states from the measurable correlation data of a quantum computer. We show that the statistics of the measured correlators contains sufficient information to characterise the entanglement, and to quantify the mixedness of the state of the computer's register. The transition to the maximally mixed regime, in the embedding space, displays a sharp boundary between entangled and separable states. Away from this boundary, the multipartite entanglement structure is robust to finite noise.

\end{abstract}

\maketitle

Entanglement is a genuine quantum feature reflected in correlations between subsystems, which persist irrespective of the locally chosen measurement bases.
Specifically, entanglement of mixed states (bi- and multipartite) is an active and important research area, both from a foundational perspective as well as for applications in quantum information processing.
The discrimination between entanglement and classical correlations is crucial to identify the genuine quantum features of a system.
It involves measurements of non-commuting observables on the entangled subsystems, for example in Bell tests \cite{aspect_experimental_1981,aspect_experimental_1982,bell_speakable_1987}, where the measurement statistics shows correlations that cannot be generated in classical systems.
These quantum correlations make multipartite entanglement a necessary (although not sufficient) resource for a potential quantum advantage over classical computation \cite{jozsa_entanglement_1997,jozsa_role_2003,shor_algorithms_1994,grover_fast_1996,deutsch_quantum_1997,wootters_quantum_1998}.
It is also known that quantum computing platforms---especially at the current ``noisy intermediate-scale quantum'' (NISQ) level---suffer from classical noise, leading to mixedness of the quantum register state, that potentially destroys the entanglement and corrupts the potential computational advantage.
Consequently, robust strategies to assess multipartite entanglement properties in the presence of such noise are indispensable.

The multipartite setting is characterised by a plethora of distinct entanglement classes and concepts \cite{dur_three_2000,sorensen_entanglement_2001,mintert_measures_2005,horodecki_quantum_2009,spee_maximally_2016}.
For large systems, for example quantum computers of sufficiently many qubits that can be used to address relevant problems, an exhaustive description of the multipartite entanglement properties is illusive.
Therefore, suitably defined observables \cite{mintert_measures_2005} and/or effective statistical characterisation tools are necessary to detect the distinctive imprint of many-body coherence and entanglement on measurable data, which has provoked substantial research interest in the recent past \cite{islam_measuring_2015,walschaers_statistical_2016,zhou_detecting_2019,knips_multipartite_2020,amaro_design_2020,amaro_scalable_2020,chen_detecting_2021,brunner_many-body_2022,wall_tensor-network_2022,ketterer_statistically_2022}.
Since the Hilbert space of multi-particle quantum systems grows exponentially with the particle number, one typically requires high-dimensional datasets---although ideally scaling only polynomially with the system size---to resolve the quantum properties of such systems  with adequate accuracy.
This constitutes a perfect application ground for machine learning algorithms, which are designed to extract relevant, problem-specific quantifiers from such high-dimensional datasets.

In our present contribution, we study \textit{entanglement partitions} of multi-qubit states, i.e. we  identify the subsets of qubits in which the quantum state separates, while qubits within these subsets are mutually entangled.
We address this problem with a data-driven ansatz. We sample the state space with randomly generated multi-qubit states of a given partition and the sampled states are characterised through correlation measurements in random measurement bases of the individual qubits.
The statistical features of such random correlator datasets, which should be accessible in any reasonably controlled quantum processing device, are indicative of the underlying entanglement partition. Our goal is to employ a non-linear dimensionality reduction algorithm (also known as manifold learning) to extract features, such as to learn how to discriminate a measured state's entanglement partition \cite{brunner_interference_2023}.
In the low-dimensional embedding space, we are further able to systematically study which correlation orders and statistical signatures of the measured distributions are the most distinctive for the identification of entanglement partitions.

Based on the proposed method, we also analyse multipartite entanglement along the transition to the (separable) maximally mixed state \cite{fine_equilibrium_2005}. For this, we introduce a novel, heuristic entanglement quantifier for a given partition, the \textit{partition-log-negativity}.
We show that the deployed algorithm identifies a clear boundary between entangled and separable states.
This cut defines reasonably large domains in which the entanglement partitions, as well as their detection via the  statistical approach  introduced here, are stable under noise.

\section{Methods: Random correlators and manifold learning}

\subsection{Entanglement partitions and measurement scheme}
\label{sec:ent_class:randomised_measurement_method}

\begin{figure}
	\centering
	\includegraphics[width=\linewidth]{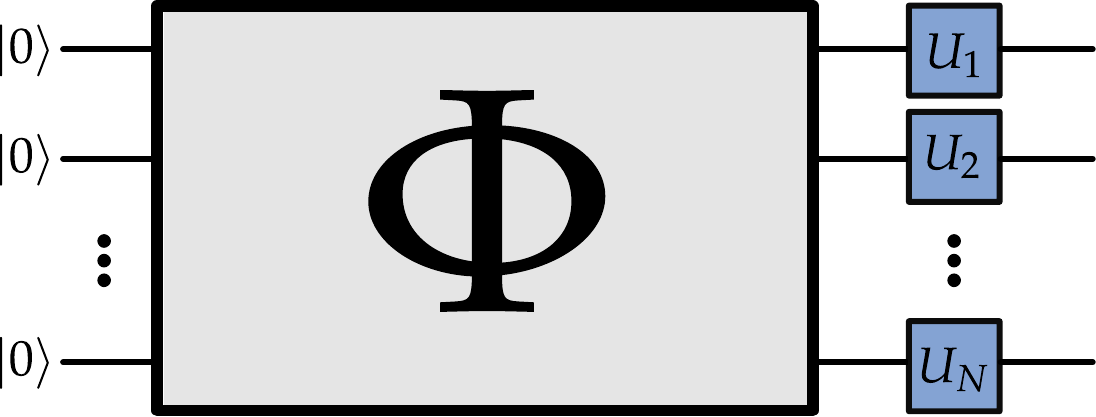}
	\caption{
		Random correlator measurement scheme.
		The unknown circuit $\Phi$ generates an entangled $N$-qubit state $\rho$.
		Independent Haar-random unitaries $U_i$ are applied to each qubit, after which correlators of order $k=1, \dots, N$ [cf. Eq.~\eqref{eq:ent_class:random_correlator}] are measured. The connected $k$-point correlators [cf. Eq.~\eqref{eq:ent_class:random_con_correlator}] obey a probability distribution which is induced by the Haar distribution of the $U_i$, and which is here characterised by its first $t_0$ statistical moments [cf. Eq.~\eqref{eq:ent_class:statistical_moments}].
	}
	\label{fig:ent_class:randomised_correlators}
\end{figure}

We consider arrays of qubits, i.e. two-level systems, due to their immediate practical relevance for quantum computation.
However, the following ideas can in principle also be applied to higher-dimensional system constituents.

We assume that a noisy quantum circuit $\Phi$  generates a mixed entangled state on output. 
To determine the state's specific entanglement structure, we study probability distributions of correlators of subsets of the constituent qubits, with each qubit being measured in a randomly chosen basis.
The basic setup is depicted in Fig.~\ref{fig:ent_class:randomised_correlators}.

The entanglement partition can be described by a partition of the qubit register, i.e. of the set $\lbrace 1,\dots, N \rbrace$.
For example, for $N = 3$, the set of such partitions is given by
\begin{equation}\label{eq:ent_class:partitions_3qubits}
	\begin{split}
		\mathbb{P} = \left\lbrace \begin{array}{l}
			\vspace{2mm} 
			[[1,2,3]],   \\
			\vspace{2mm} 
			[[1],[2,3]], [[1,3],[2]],  [[1,2],[3]],  \\
			\vspace{2mm}  
			[[1],[2],[3]]
		\end{array} \right\rbrace
	\end{split} \,.
\end{equation}
Qubits within one part $P_i$ of the partition $P = [P_1,\dots, P_r] \in \mathbb{P}$ are entangled, qubits belonging to different parts are not.
In the example \eqref{eq:ent_class:partitions_3qubits}, $P = [[1], [2,3]]$ describes three qubits where qubit two and three are entangled ($P_2 = [2,3]$), while the first qubit ($P_1 = [1]$) separates from them. The number of possible partitions is called \textit{Bell number} and grows extremely fast with $N$\cite{bell_exponential_1934}.
We denote by $|P|$ and $|P_i|$ the number of parts of $P$ and the number of qubits in $P_i$, respectively. 

A pure state with entanglement partition $P$ is given by
\begin{equation}\label{eq:ent_class:random_pure_part_state}
	\ket*{\psi} = \ket*{\psi_{P_1}} \otimes \dots \otimes \ket*{\psi_{P_r}} \,,
\end{equation}
where each factor $\ket*{\psi_{P_i}}$ is an entangled state of the qubits in $P_i$.
To simulate the output of the noisy state generation circuit $\Phi$ [cf. Fig.~\ref{fig:ent_class:randomised_correlators}], we consider random mixtures of states of type \eqref{eq:ent_class:random_pure_part_state}.
For $P \in \mathbb{P}$ we generate $n_\mathrm{mixed}$ such pure states, where each $\ket*{\psi_{P_j}}$ is drawn at random from the Haar measure on the $|P_j|$-qubit space.
Additionally, random real numbers $c_1,\dots, c_{n_\mathrm{mixed}}$ are sampled uniformly between zero and one, and then renormalised such that $\sum_i c_i = 1$.
This defines a random mixed state with entanglement partition $P$:
\begin{equation}\label{eq:ent_class:random_part_state}
\rho  = \sum_{i=1}^{n_\mathrm{mixed}} c_i \op*{\psi^i} \,.
\end{equation}
The states generated in this way sample the quantum state space of mixed entangled states with the given partition. They are used to construct a dataset (via the measurement protocol introduced below) with which we train a state-agnostic classification model to identify the entanglement partitions of general mixed states.
In the following, if not stated differently, we set $n_\mathrm{mixed} = 10$.
This choice serves as a proof-of-concept for the introduced protocol (which performs similarly for other choices of $n_\mathrm{mixed}$).
By increasing $n_\mathrm{mixed}$, we could probabilistically reduce the purity of the states \eqref{eq:ent_class:random_part_state}.
We leave the exploration of this parameter for further research.
Another obvious alternative to introduce unbiased mixedness is via a global depolarizing channel.
This approach will be employed in Sec.~\ref{sec:ent_class:transition_to_mixed} to systematically study the entanglement structure of states \eqref{eq:ent_class:random_part_state} along a continuous path towards the (separable) maximally mixed state.
Note, however, that the entanglement partition of a given state is, in general, destroyed under a global depolarising channel.

At the output of $\Phi$, we measure correlations in the computational basis after rotating the qubits by independent random Haar unitaries $U_1,\dots, U_N$. This pre-measurement random rotation should not to be confused with the random generation of states according to Eq.~\eqref{eq:ent_class:random_part_state} (using the Haar measure on the individual parts of $P$ and not on the single qubits).
A $k$-point correlator on qubits $m_1,\dots, m_k$ is given by
\begin{equation}\label{eq:ent_class:random_correlator}
c_{m_1,\dots, m_k} = \expv{U_{m_1}^\dagger Z_{m_1} U_{m_1} \dots U_{m_k}^\dagger Z_{m_k} U_{m_k}} \,,
\end{equation}
where $Z_m$ is the Pauli-$Z$ operator on qubit $m$ and the expectation value is taken in the state $\rho$ generated by the noisy quantum process $\Phi$.
All correlators of order up to $k$ contribute to the connected (also known as truncated) correlators
\begin{equation}\label{eq:ent_class:random_con_correlator}	\mathcal{C}_{m_1,\dots, m_k} = \sum_{P \vdash \lbrace m_1, \dots, m_k \rbrace } (|P| - 1)! (-1)^{|P| - 1} \prod_{P_i \in P}
c_{P_i} \,,
\end{equation}
where the symbol $\vdash$ denotes partitions $P$ of the set $\lbrace m_1, \dots, m_k \rbrace$. The product runs over all parts $P_i$ of $P$ and $c_{P_i}$ is the correlator \eqref{eq:ent_class:random_correlator} of the qubits in $P_i$.
For example, for $k=2$ we have $\mathcal{C}_{m_1, m_2} = c_{m_1, m_2} - c_{m_1} c_{m_2}$, which is the covariance.
In the mathematical literature, the connected correlators are also known as \textit{joint cumulants}. They appear in different disciplines, such as statistics and the theory of large deviations 
\cite{varadhan_asymptotic_1966,varadhan_large_1984,klenke_probability_2008}, or in quantum field theory in the form of connected Green's functions
\cite{feynman_theory_1949,schwartz_quantum_2013}.

Based on the distributions of the (connected) correlators, we wish to characterise entanglement partition and mixedness of the quantum states generated by $\Phi$.
Formally, the set of correlators up to order $k_0$ contains the same information as the set of connected correlators up to order $k_0$.
However, the $\mathcal{C}_{m_1,\dots, m_k}$ directly probe the factorisation properties of the $c_{m_1,\dots, m_k}$, i.e. $\mathcal{C}_{m_1,\dots, m_k}$ vanishes if a single qubit is uncorrelated from all other qubits.  
There are $\sum_{k=1}^{k_0} \binom{k_0}{k} = 2^{k_0} - 1$ connected $k$-point correlators with $k \leq k_0$.
With respect to the Haar distribution on each qubit, each correlator exhibits a probability distribution which we condense into the finite set of its first $t_0$ statistical moments
\begin{equation}\label{eq:ent_class:statistical_moments}
	M^{t}_{m_1, \dots, m_k} = \int dU_1 \dots dU_k \left( \mathcal{C}_{m_1, \dots, m_k} \right)^t \,, 
\end{equation}
for $t = 1,\dots, t_0$ \cite{brunner_interference_2023}.
A similar approach was followed in \cite{brunner_many-body_2022} to quantify the coherence of a many-body (e.g. photonic) state from the first statistical moment of correlators at the output of a non-interacting Haar unitary.
Note that, because $c_{m_1, \dots, m_k} \rightarrow - c_{m_1, \dots, m_k}$ under reflection at one of the local Bloch spheres \cite{ketterer_statistically_2022}, all odd moments of $c_{m_1, \dots, m_k}$ are zero, independently of the measured states $\rho$.
This is not necessarily the case for the odd moments of $\mathcal{C}_{m_1, \dots, m_k}$. However, we expect that the odd moments of the connected correlators carry less information about the measured states, such that we disregard them in the following analysis.  
The dimension of the data vector containing the even statistical moments up to $t_0$ of random correlators up to order $k_0$, for a given random state \eqref{eq:ent_class:random_part_state}, is
\begin{equation}\label{eq:ent_class:correlator_data_size}
	D 
 = \frac{t_0}{2} \big( 2^{k_0} -1 \big) \,.
\end{equation}
In our following analyses, we will mostly truncate this data vector to a smaller dimensionality ($D \leq t_0(2^{k_0} -1)/2$), by taking into account only specific statistical moments or correlation orders.

In practice, the moments $M^t_{m_1,\dots, m_k}$ can only be retrieved from finite statistics with limited accuracy, by sampling over the independent, random local unitaries $U_i$.
We include this statistical uncertainty in our algorithm training (discussed below) by calculating the $M^t_{m_1,\dots, m_k}$ from a finite number of $k$-tuples $(U_{m_1}, \dots, U_{m_k})$ of independent random unitaries.
For an in depth analysis of the statistical error incurred by this procedure, we refer the reader to \cite{ketterer_statistically_2022}.
There, the authors study concentration bounds, such as Chebyshev-Cantelli or Bernstein inequalities, to derive the optimal numbers of distinct $k$-tuples of random unitaries $U_i$, and of measurement repetitions for each set of unitaries, to estimate the moments \eqref{eq:ent_class:statistical_moments} with accuracy $\delta$ and confidence $\gamma$, i.e.
\begin{equation}
	\mathrm{Prob} \left( \left| \tilde M_{m_1, \dots m_k}^t - M_{m_1, \dots m_k}^t \right| \leq \delta \right) \geq \gamma \,,
\end{equation}
where $\tilde M_{m_1, \dots m_k}^t$ is the estimate of the true moment inferred from the finite statistics.
For the second moment, $t=2$, which will be the most relevant for our following investigations, they find that, for given correlation order $k$ and confidence $\gamma$, the accuracy $\delta$ scales as $\sim 1/\sqrt{N_\mathrm{unit}}$, where $N_\mathrm{unit}$ is the number of random unitary $k$-tuples $(U_{m_1}, \dots, U_{m_k})$.
In turn, the number $N_\mathrm{unit}$ required to achieve a given accuracy $\delta$ and confidence $\gamma$ scales exponentially in the correlation order $k$
\footnote{
Note that, in \cite{ketterer_statistically_2022}, the ordinary correlators [Eq. \eqref{eq:ent_class:random_correlator}] were considered, rather than the connected correlators [Eq. \eqref{eq:ent_class:random_con_correlator}].
However, we can estimate, for given correlation order $k$, all $\ell$-point correlators with $\ell < k$ (and products thereof)---required to build the connected correlator---from the same measurement statistics. Since the number $N_\mathrm{unit}$ of unitaries required for a given accuracy $\delta$ grows with $k$, the moment estimates of lower order correlators are more accurate than for the $k$-th order term. Therefore, the estimation of statistical moments of the connected correlators and of the ordinary ones exhibits the same scaling behaviour.}.
Hence, analysing random correlator datasets containing only relatively low-order correlators is of particular interest to obtain a scalable protocol.
Below, in Sec.~\ref{sec:ent_class:comparing_correlators}, we will specifically address this point by analysing the capabilities of the individual correlation orders to discriminate between various entanglement partitions. 
In general, one can obtain an optimal decomposition of the total number of measurements into $N_\mathrm{unit}$ random choices of measurement bases and a certain number of measurements per basis, which, however, strongly depends on the required accuracy and confidence \cite{ketterer_statistically_2022}.
For our proof-of-principle numerical studies we will always estimate Eq.~\eqref{eq:ent_class:statistical_moments} from $N_\mathrm{unit} = 500$ unitary $k$-tuples $(U_1,\dots, U_k)$.

The statistical moments of the random correlators should be readily accessible with high accuracy in state-of-the-art quantum hardware, since single-qubit rotations and readout are the most basic building blocks for any quantum processing task.
The distributions of the random correlators and their statistical moments \eqref{eq:ent_class:statistical_moments} have found a wide range of applications for characterizing coherence and entanglement in quantum systems: For example the well-established Bell tests rely on the a statistical analysis of such correlations in random measurement bases \cite{aspect_experimental_1981,aspect_experimental_1982,bell_speakable_1987}.
Random two-point correlators where employed in \cite{walschaers_statistical_2016,giordani_experimental_2018,brunner_many-body_2022} to quantify many-particle coherence and particle (in)distinguishability in non-interacting (e.g. photonic) many-body systems.
Moreover, in \cite{van_enk_measuring_2012,elben_statistical_2019} the authors showed that the moments of the full measurement statistics give access to purity and Rényi entropies of the measured quantum state.
In \cite{ketterer_statistically_2022,ohnemus_quantifying_2023}, a detailed statistical analysis of the moments (for the non-connected correlators) was carried out, from which mixed-state witnesses of $k$-separability were derived.
Furthermore, \cite{knips_multipartite_2020} considered directly the full distributions of the random correlators \eqref{eq:ent_class:random_correlator}. They showed that comparing those with the distributions of lower-order factorisations of the correlators allows to detect multipartite entanglement of certain GHZ states or cluster states.

Our aim is to bypass such comparison by hand and to feed the random correlation data directly into a machine-learning model.
In this way we let the algorithm figure out how to combine the various correlation orders and moments in such a way as to extract the most discriminative features \cite{giordani_experimental_2018} of the state's entanglement partition.
This also allows us to systematically study
possible truncations of the dataset---for example including only low statistical moments or  correlation orders---and how this truncation reduces the discriminative power of the dataset.
These questions are important for applications, since in actual experiments the limited measurement statistics only allows to accurately estimate low-order statistical quantifiers.

\begin{figure*}
	\includegraphics[width=\linewidth]{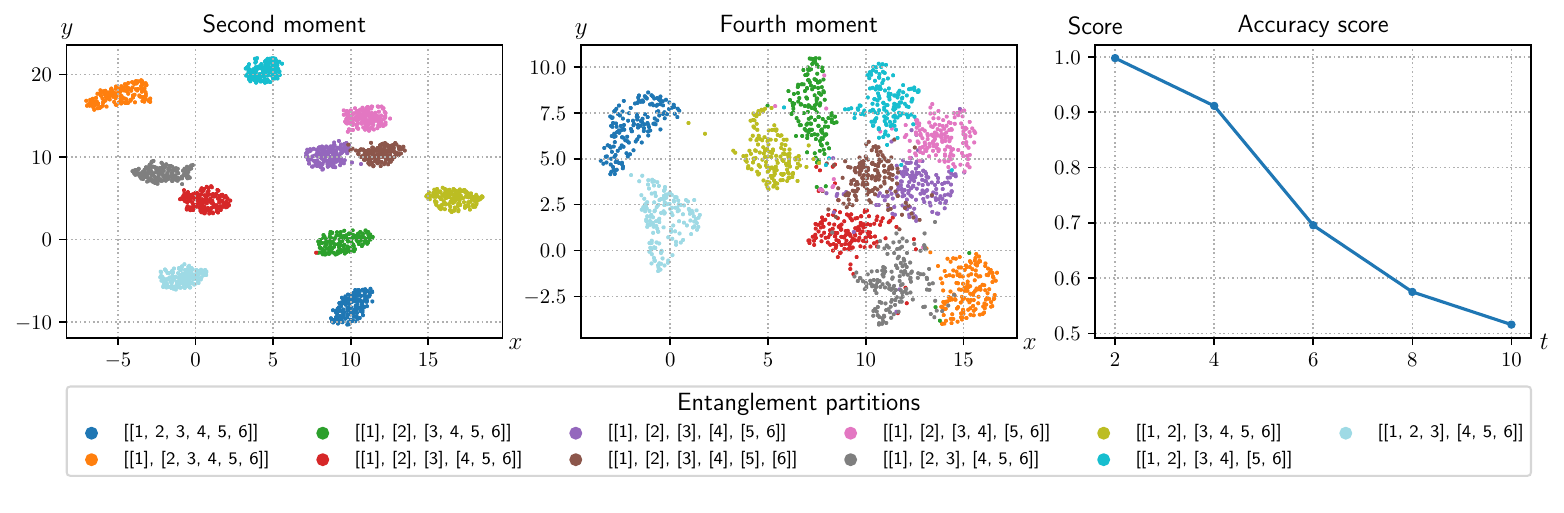}
	\caption{
		Comparison of the individual statistical moments $M^t_{m_1,\dots, m_k}$, $t=2,4,\dots, 10$, for $N=6$ qubits. The left and middle panel show the UMAP embedding of the data of the second ($t=2$) and fourth ($t=4$) statistical moments.
		The points are coloured according to their entanglement partitions, which are given in the legend.
		We only consider ordered partitions, since for $N=6$ there are already $203$ partitions in total [cf. Fig.~\ref{fig:ent_class:all_partitions}].
		For $t=2$, the emergence of well-separated clusters for each partition can be observed.
		Note that the fully separable partition $[[1],[2],[3],[4],[5],[6]]$ and $[[1],[2],[3],[4],[5, 6]]$ are embedded relatively close to each other, due to their similarity.
		The clusters partially overlap and get more diffuse for $t=4$.
		For a quantitative picture, the right panel shows the accuracy score \eqref{eq:ent_class:score}, inferred from classifying unseen test data with the trained classification model, as a function of $t$.
		Almost perfect classification is achieved for $t=2$. For increasing $t$, the score monotonically decreases.}
	\label{fig:ent_class:moment_comparison}
\end{figure*}

\subsection{Manifold learning and classification pipeline}\label{sec:ent_class:manifoldlearning}

The dimension [cf. Eq.~\eqref{eq:ent_class:correlator_data_size}] of data vectors obtained from large quantum systems, such as the proposed random correlator dataset, often scales exponentially in the particle number---or in our case in the maximal correlation order $k_0$.
Such a high-dimensional dataset potentially contains information that is redundant or irrelevant for the task at hand.
Machine learning can be used to condense this information into a smaller set of relevant features, i.e. into lower-dimensional representations (whose dimension, for example, scales only polynomially or even is constant in the particle number).
An often made assumption is that the relevant features or variables define a low-dimensional manifold (i.e. a curved space), which is embedded in the Euclidean (i.e. with a flat Riemannian metric) data space $\mathbb{R}^D$ (with $D$ given e.g. by Eq.~\eqref{eq:ent_class:correlator_data_size}).
Manifold learning, also known as non-linear dimensionality reduction, tries to identify this underlying geometric structure and to generate a faithful low-dimensional representation thereof \cite{lee_nonlinear_2007}.
Common algorithms are, for example, isomap \cite{tenenbaum_global_2000}, locally linear embedding \cite{roweis_nonlinear_2000}, autoencoder \cite{kramer_nonlinear_1991}, or $t$-SNE \cite{hinton_stochastic_2002,maaten_visualizing_2008}.

In this contribution we apply UMAP, the \textit{uniform manifold approximation and projection} algorithm, introduced in \cite{mcinnes_umap_2020}, which has already been applied for feature extraction and visualisation in many fields such as biology, machine learning or time-series analysis \cite{cao_single-cell_2019,diaz-papkovich_umap_2019,carter_activation_2019,ali_timecluster_2019}.
The basic idea behind this and also other manifold learning methods is as follows:
Consider data points $x_i, i = 1,\dots, n_\mathrm{data}$ in some (potentially high-dimensional) vector space $X = \mathbb{R}^D$ of dimension $D$.
In our case, $x_i = (x_i^1,\dots, x_i^D)$ is the vector of statistical moments [Eq.~\eqref{eq:ent_class:statistical_moments}] of the random $k$-point correlators [Eq.~\eqref{eq:ent_class:random_con_correlator}], or a sub-vector thereof containing only specific statistical moments or correlation orders.
We wish to embed this dataset via an injective mapping $f : X \rightarrow Y$ into a lower-dimensional space.
To find $f$, a weighted graph $G_X$ is generated on the points $x_i$.
The edge weights  $v_{ij} = g_{x_j}(\mathrm{dist}_X(x_i, x_j) )$ are determined by a similarity function $g_{x_j}$ which associates large (small) values to pairs $x_i, x_j$ which are close together (far away) in terms of the standard Euclidean distance in $X=\mathbb{R}^D$.
The function $g_{x_j}$ can explicitly depend on the base point $x_j$ (or its local neighbourhood).
Usually the underlying manifold is not covered uniformly by the data with respect to the Euclidean metric. Because of this, UMAP defines $g_{x_j} \propto \exp\left( -(\mathrm{dist}(x_i, x_j) - \rho_j) / \sigma_j \right)$, where $\rho_j$
is the minimal distance between $x_j$ and any other point $x_i$ and $\sigma_i$ is a width parameter, defined in \cite{mcinnes_umap_2020}.
If the neighbourhood of $x_j$ is only sparsely populated, $\rho_j$ is large and $g_{x_j}$ associates comparably larger values (similarities) to pairs $x_i, x_j$ as it would in a denser region. In this way, a $x_j$-dependent metric is induced with respect to which the data covers the manifold approximately uniformly \cite{mcinnes_umap_2020}.
The edge weights $v_{ij}$ are interpreted as a probability distribution over $G_X$: large values of $v_{ij}$, typically associated to small distances $\mathrm{dist}_X(x_i, x_j)$, correspond to a large probability that the points $x_i$ and $x_j$ are `connected', and vice versa.

To each data point $x_i\in X$, one associates a point $y_i\in Y$ in the embedding space. One then defines a graph $G_Y$ on the embedded points $y_i$, with edge weights $w_{ij} = h_{y_j}(\mathrm{dist}_Y(y_i, y_j))$ between points $y_i$ and $y_j$. As above, $\mathrm{dist}_Y(y_i, y_j)$ is a distance in $Y$ and $h_{y_j}$ is a similarity function.
The positions of the embedded points $y_i, y_j$ are then optimised in such a way as to minimise a distance measure between the distributions $v_{ij}$ and $w_{ij}$,
for example the Kullback-Leibler divergence \cite{kullback_information_1951}, or a variation thereof.
The embedding is the result of this optimization.
Often used embedding spaces are $Y = \mathbb{R}^2$ or $Y = \mathbb{R}^3$, since these allow for a direct visualisation of the embedded data and its geometry.

The precise definitions of metrics and similarity functions of UMAP can be found in \cite{mcinnes_umap_2020}.
There, also its hyperparameters are introduced in more detail. Most importantly, one needs to adjust three parameters: The dimension of the embedding space $Y$, the number $n_\mathrm{neighbours}$ of nearest neighbours which are taken into account to calculate the edge-weights $v_{ij}, w_{ij}$ on each point in the graphs, and $d_\mathrm{min}$, which sets a minimal allowed distance between points in the low-dimensional embedding.
The interplay of $n_\mathrm{neighbours}$ and $d_\mathrm{min}$ controls a trade-off between preserving global versus local structures in the embedding \cite{mcinnes_umap_2020}.
More precisely, small values of $n_\mathrm{neighbours}$ and $d_\mathrm{min}$ typically result in the formation of many isolated clusters, i.e. the algorithm groups together only data points with a high degree of similarity, as measured by the weights $v_{ij}$. In turn, large values of $n_\mathrm{neighbours}$ and $d_\mathrm{min}$ try to prevent the embedded data from splitting into clusters.
For the embedding dimension, we always use a two-dimensional space $Y = \mathbb{R}^2$ in the following.

To quantify the ability of UMAP to identify the entanglement partition of given input states, we apply a classifier on the embedded data.
We use a standard decision tree classifier of the \textit{scikit-learn} python package \footnote{The decision tree (a commonly used classification algorithm) aims to identify a piecewise constant approximation to the data-label pairs by extracting simple decision rules from the data \cite{hastie_elements_2009}.}.
In this sense, UMAP can be seen as a feature extraction step of a classification pipeline to characterise entanglement partition and mixedness of the measured quantum states $\rho$.
In a first step, before UMAP is applied, we standardise our correlator dataset to zero mean and unit variance by a linear transformation. This is necessary, since common feature extraction and classification algorithms are typically optimised to such standardised data.
The full pipeline defines a mapping
\begin{equation}
	F : X \; \stackrel{\text{standardisation}}{\longrightarrow} \; \tilde X \; \stackrel{\text{feature extraction}}{\longrightarrow} \; Y \; \stackrel{\text{classification}}{\longrightarrow} \; \mathbb{P} \,,
\end{equation}
from the data space $X$ to the set of partitions $\mathbb{P}$ of $N$ qubits [cf. Eq.~\eqref{eq:ent_class:partitions_3qubits}].
The final classification step needs to be trained in a supervised manner.
For this we feed the full pipeline with a training dataset consisting of input-output pairs $(x_i, P^{(i)}), i=1,\dots, n_\mathrm{data}$, with $x_i = (x_i^1,\dots, x_i^D) \in X$  the data of statistical moments \eqref{eq:ent_class:statistical_moments} and $P^{(i)}$  the  partition of the measured state.
During training, the decision tree is given pairs $(y_i, P^{(i)})$ of the UMAP-embedded points $y_i$ with the corresponding entanglement partition $P^{(i)}$. Note that the UMAP dimensionality reduction is \textit{unsupervised}, i.e. the algorithm only accesses the $x$-component (the random correlator data) of the tuples $(x, P)$.
After training, we test the pipeline with a newly generated (unseen by the algorithm) test dataset, typically of the size of around $10\%$ of the size $n_\mathrm{data}$ of the training dataset.
The $x$-component of the test tuples $(x,P)$ is mapped via $F$ to a partition $P'$. This is compared to the actual $P$-component of the training tuple.
To quantify the performance of the classification we consider the ratio of the number  $n_\mathrm{correct}$ of correct classifications to the total number $n_\mathrm{correct} + n_\mathrm{incorrect}$ of test samples, referred to as accuracy score
\begin{equation}\label{eq:ent_class:score}
	\mathrm{score} = \frac{n_\mathrm{correct}}{n_\mathrm{correct} + n_\mathrm{incorrect}} \,,
\end{equation}
with $n_\mathrm{incorrect}$ the number of false classifications. 

\section{Results}

In the following, we employ the above classification scheme to systematically study the information about a given state's entanglement partition and degree of mixedness contained in the first $t_0$ statistical moments of the random connected correlators of order $1,\dots, k$---which define entries of our $D$-dimensional data points in $X$.

\subsection{Comparing the statistical moments}
\label{sec:compare_moments}

We start by comparing the different moments of the correlators.
We consider a system of $N = 6$ qubits and take all correlation orders into account, i.e. we set $k=N$.
For each partition $P\in \mathbb{P}$ we generate $200$ random states $\rho$ given by Eq.~\eqref{eq:ent_class:random_part_state} with $n_\mathrm{mixed} = 10$.
For illustrative purposes and since there are already $203$ partitions for six qubits, we at first only consider the $11$ \textit{ordered partitions}, i.e. those $P = [P_1, \dots, P_r]$ where both the lengths of the parts $P_i$ and the qubit indices increase from left to right [i.e. the first elements of each line in Eq.~\eqref{eq:ent_class:partitions_3qubits}].
For each of the resulting $n_\mathrm{data} = 11 * 200$ states, we calculate a corresponding $D$-dimensional data vector with $t_0 = 10$ [cf. Eq.~\eqref{eq:ent_class:correlator_data_size}].

To resolve the discriminative power of individual moments, we separately train the UMAP-based classification pipeline on sub-datasets which only contain a single statistical moment $t=2,4,6,8,10$, respectively.
The left and middle panels of Fig.~\ref{fig:ent_class:moment_comparison} display the embedded data for $t=2$ and $t=4$.
We choose relatively small hyperparameters $n_\mathrm{neighbours} = 10$ and $d_\mathrm{min} = 0.6$, favouring the formation of many independent clusters. This is a suitable choice for this setting, where we seek to classify the data with respect to a large number of classes, i.e. the different entanglement partitions
\footnote{We tested various parameter combinations and numerically observed, except for very large ($\sim 10^3$) and very small ($\sim 1$) numbers $n_\mathrm{neighbours}$, qualitatively very similar embeddings.}.
In the figure, the embedded points are coloured according to their entanglement partition [Eq.~\eqref{eq:ent_class:partitions_3qubits}]---which is not disclosed to UMAP---as indicated in the legend.

For $t = 2$, we see that clearly resolved clusters emerge for different entanglement partitions.
Analogously generated clusters partially overlap for $t=4$. 
From this visual impression, it is clear that the discrimination task is better solved when assessing the data of second moments.
To quantify the performance of the classification of the embedded data, we plot the accuracy scores \eqref{eq:ent_class:score} achieved on the basis of the (even) moments $t \leq 10$ (for an unseen test dataset of 20 states per partition and containing only the respective moment) in the right panel of Fig.~\ref{fig:ent_class:moment_comparison}.
The score monotonically decreases with increasing $t$.
For $t=2$ the score is almost one, confirming a very accurate classification of entanglement partitions based on the second moments of the random correlators.
The fourth moment still performs reasonably well, achieving a score of above $0.9$.
Note that while the embedded points associated with distinct partitions occasionally also exhibit distinct 2D shapes, the precise geometric features and the coordinates of the embedded data do not play a role for the classification task---only the appearance of clusters is relevant.

\begin{figure}
	\includegraphics[width=\linewidth]{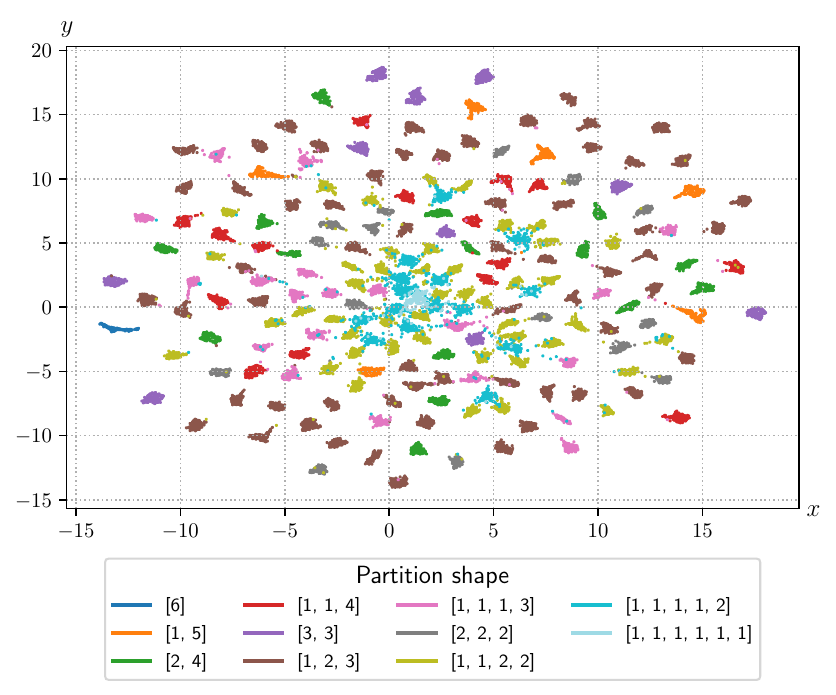}
	\caption{
		Embedding \textit{all} $203$ entanglement partitions for $N=6$ qubits, based on the second moments ($t = 2$) of the random correlators [Eq.~\eqref{eq:ent_class:statistical_moments}].
		The partitions are grouped and coloured according to their \textit{shapes} [explained in the main text], which are given in the legend.
		The embedded data points from each of the $203$ partitions form clear clusters.
		The more separable partitions (with shape $[1,1,1,1,2]$ and $[1,1,2,2]$) are located in the middle of the plot, around the fully separable one $[1,1,1,1,1,1]$.
		Based on this UMAP embedding, we obtain an accuracy score \eqref{eq:ent_class:score} of $0.93$ for the classification of all partitions, which certifies the suitability of the proposed scheme for this large-scale classification task.}
	\label{fig:ent_class:all_partitions}
\end{figure}

Based on the second moment [cf. Eq.~\eqref{eq:ent_class:statistical_moments}] of the random correlator dataset, we demonstrate the capabilities of the proposed pipeline to discriminate between \textit{all} $203$ partitions of a $N = 6$ qubit system.
For each partition, we generate $200$ random states with $n_\mathrm{mixed} = 10$ [cf. Eq.~\eqref{eq:ent_class:random_part_state}].
The embedded data derived from the second moment of random correlators is shown in Fig.~\ref{fig:ent_class:all_partitions}.
As above, we set $n_\mathrm{neighbours} = 10$ and $d_\mathrm{min} = 0.6$.
The points are coloured according to the \textit{partition shape}, i.e. the ordered tuple of lengths $|P_i|$ of parts in $P$ (given in the legend).
For all partitions, well-separated clusters appear.
For a quantitative picture we, again, calculate the accuracy score achieved in the classification of a test dataset. Over all partitions, an accuracy score of $0.93$ is achieved. This clearly demonstrates the applicability of the proposed scheme to also capture the fine differences between entanglement partitions of the same shape.
The trained model can be used to embed new data obtained from measurements on actual quantum devices, which yields an efficient classification tool to identify the entanglement partition of states generated by such devices.
In \cite{new_paper_inprep}, the presently developed protocol is applied to certify entanglement partitions of generalized GHZ and W states prepared on IBM quantum devices. Our method is able to correctly identify the entanglement partitions of these states, with almost perfect accuracy.

Although this is not directly relevant for the classification problem, we observe that partitions (e.g. $[[1],[2],[3,4,5,6]]$ and $[3],[4],[1,2,5,6]$) belonging to the same shape (e.g. $[1,1,4]$) are spatially arranged in a specific way.
For example, for shapes $[1,1,2,2]$ and $[1,1,1,1,2]$ (light green and cyan), all associated clusters surround the fully separable partition $[1,1,1,1,1,1]$ (light blue) in the middle of the plot.
The remaining partitions are organised around this centre, in a ring-like structure.

\subsection{Comparing correlation orders}
\label{sec:ent_class:comparing_correlators}

\begin{figure}
	\includegraphics[width=\linewidth]{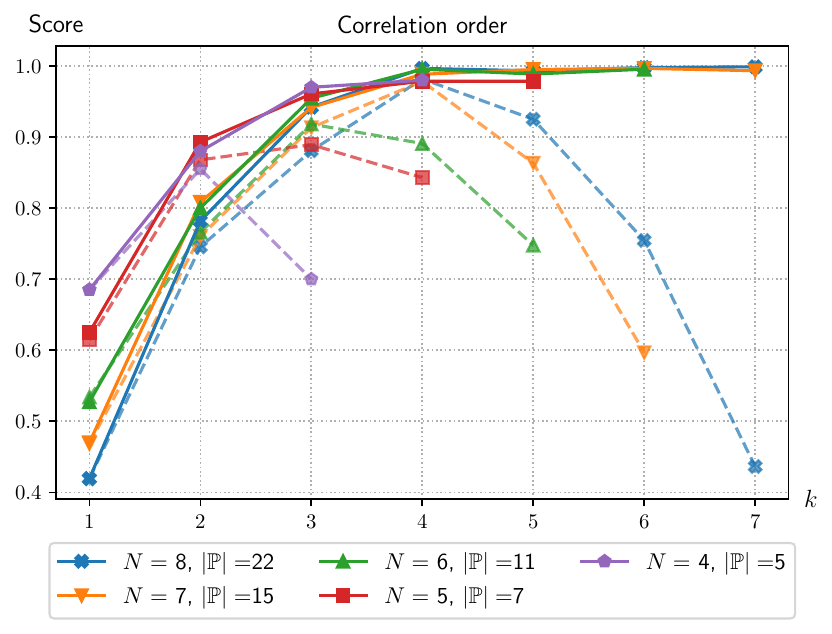}
	\caption{
		Comparison of the distinctive power of the individual correlation orders $k$.
		The accuracy score \eqref{eq:ent_class:score} is shown for the classification of ordered entanglement partitions based on data containing only correlators of specific orders, for systems of $N=4$ to $N=8$ qubits (coloured according to the legend).
		We consider two cases: Solid lines indicate the classification performance when the data of all correlators of orders $\ell \leq k$ are included. Dashed lines show the accuracy score when we include correlators exclusively of order $k$. In this case, the best classification performance is obtained from the correlation order $k \approx N/2$. For this optimal correlation order and $N \geq 6$, an accurate classification with scores above $0.9$ is achieved.
		As expected, the case where correlators up to order $k$ are considered (solid lines) consistently shows larger accuracy scores as compared to the case where only the individual correlation orders are included. Furthermore, and also as expected, the solid curves monotonically increase with $k$, since the information content of the dataset is increased for each added order.
	}
	\label{fig:ent_class:correlation_order_comparison}
\end{figure}

In the following, we study in more detail the discriminative power of the the individual correlation orders $k$.
Based on the results of Fig.~\ref{fig:ent_class:moment_comparison}, we restrict the data to the statistical moment $t = 2$.
For each correlation order $k$, we separately train the classification pipeline based on a UMAP embedding.
For each individually trained model, we evaluate the accuracy score \eqref{eq:ent_class:score} achieved on newly generated test data.
We perform this analysis for an increasing number of qubits, from $N = 4$ to $N = 8$. See \cite{brunner_interference_2023} for an extension of this study to also resolve the influence of entanglement depth (i.e. the length of the largest part of a given partition) on the discrimination performance.
As before, we only consider ordered partitions (the number of all partitions for $N =7,8$ is already $877$ and $4140$), and generate for each of them 200 random states as given in Eq.~\eqref{eq:ent_class:random_part_state}, with $n_\mathrm{mixed} = 10$.
We use the same hyperparameters as in Sec.~\ref{sec:compare_moments}.
Figure~\ref{fig:ent_class:correlation_order_comparison} shows the accuracy score as a function of $k$, for $N=4,5,6,7,8$ (distinguished by the colour code in the legend).
The legend also shows the number $| \mathbb{P}|$ of \emph{ordered} partitions, for each $N$.
We distinguish two cases: we either exclusively consider correlators of order $k$ (dashed lines) or we consider all correlators of order $\ell \leq k$ (solid lines).
As expected, the solid lines---including all correlators up to a specific order $k$---consistently show improved classification scores as compared to the dashed lines.
Also note that including two-point correlators and one-point spin measurements already yields a reasonable classification accuracy of around $0.8$ or higher for all considered $N$.
From the dashed lines---obtained from correlators of order $k$ exclusively---we observe the best classification performance for $k\sim N/2$.
For this correlation order, the score is above $0.8$ and close to one for larger qubit numbers $N=7,8$.
The information content of a single correlator's statistics on the measured state's entanglement partition then decreases again for orders $k > N/2$.
Interestingly, even though the number $|\mathbb{P}|$ of partitions which the algorithm needs to distinguish increases with $N$ (see legend), the classification scores of the best performing correlation order $k\approx N/2$ (i.e. the maxima of the dashed curves) increase with $N$.
This is a promising observation regarding the application of our proposed scheme to larger quantum systems.
The increase of accuracy for larger systems may be related to a concentration of measure phenomenon, in the sense that the set of random states belonging to a given entanglement partition becomes more concentrated in the exponentially large Hilbert space for increasing $N$. Similar observations where made in the context of entanglement and many-particle interference in \cite{tiersch_benchmarks_2009,tiersch_universality_2013,walschaers_many-particle_2016}.

\subsection{Transition to the maximally mixed state}
\label{sec:ent_class:transition_to_mixed}

In the preceding studies, we considered states obtained as a statistical mixture of $n_\mathrm{mixed} = 10$ pure states of a given entanglement partition [cf. Eq.~\eqref{eq:ent_class:random_part_state}].
This mixedness is introduced to represent the noise generated by the state generation circuit $\Phi$ [cf. Fig.~\ref{fig:ent_class:randomised_correlators}].
Below, we investigate the continuous crossover between pure entangled states of a given entanglement partition $P$ and the maximally mixed state (which belongs to the partition $[1,\dots, 1]$) in the UMAP embedding plane.
We show that the embedding is sensitive to the level of mixedness. Furthermore, we find that above a finite distance from the maximally mixed state, all states are entangled, as expected from topological considerations \cite{zyczkowski_volume_1998,fine_equilibrium_2005}.

\begin{figure*}
	\includegraphics[width=\linewidth]{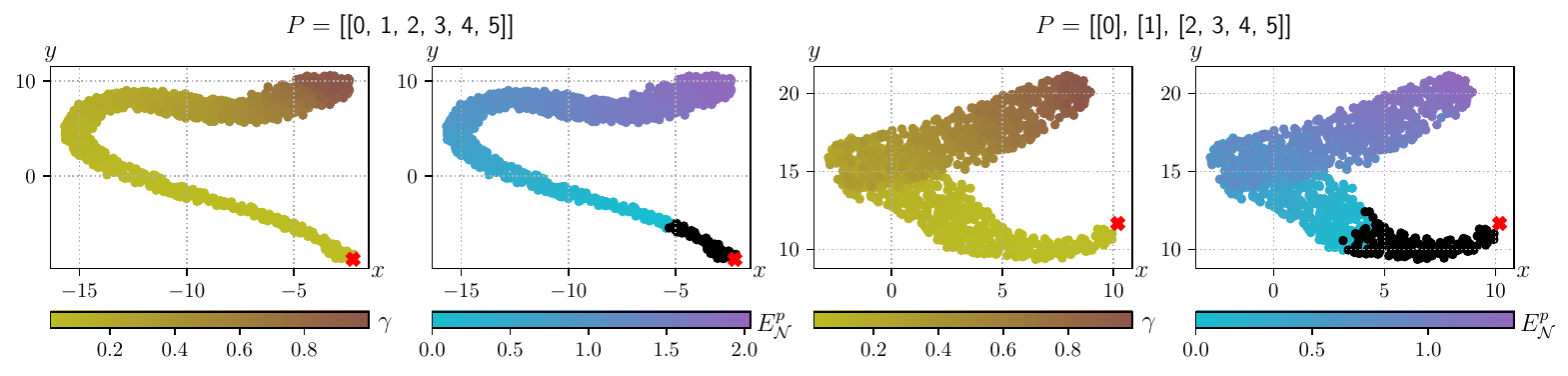}
	\caption{
		Transition to the maximally mixed state. The panels show the UMAP embeddings of the correlation data for states \eqref{eq:ent_class:mixedness_transition_state} with variable $\lambda \in [0,1]$ and for the two indicated example partitions. We consider a system of $N=6$ qubits.
		In the first and third panel, we colour the points according to the purity $\gamma$ of the measured state $\rho$. The purity is monotonically increasing along the serpentine shape of the embedding. The maximally mixed state is indicated by a red cross. It is correctly situated at the strongly mixed limit of the serpentine.
		The second and fourth panel show the same data, but coloured according to the partition-log-negativity $E^P_\mathcal{N}$ [cf. Eq.~\eqref{eq:ent_class:partition_log_negativity}], which we propose as a heuristic, mixed-state multipartite entanglement quantifier for given partition $P$.
		States with $E^P_\mathcal{N}(\rho) = 0$, which we call PPT states \cite{bengtsson_geometry_2006}, are coloured in black. There is a clear border between NPT states, $E^P_\mathcal{N}(\rho) > 0$, and PPT-states in the embedding. NPT states are genuinely entangled according to the partition $P$ [cf. discussion after Eq.~\eqref{eq:ent_class:partition_log_negativity}]. Above this cut, the partition-log-negativity increases approximately monotonically along the serpentine shape with increasing purity of $\rho$.
	}
	\label{fig:ent_class:mixednetss_transition}
\end{figure*}

We consider convex combinations
\begin{equation}\label{eq:ent_class:mixedness_transition_state}
	\rho = ( 1 - \lambda) \op{\psi} + \lambda \id_d / d \,,
\end{equation}
of random pure states $\ket{\psi}$ of a given entanglement partition $P$ [cf. Eq.~\eqref{eq:ent_class:random_pure_part_state}] with the maximally mixed state $\id_d / d$, where $d$ is the dimension of the $N$-qubit space.
We fix the partition $P$ and discretise the interval $[0,1]$ in $1000$ evenly spaced points $\lambda_1,\dots, \lambda_{1000}$. For each $\lambda_i$ we calculate a random mixed state $\rho_i$ given by Eq.~\eqref{eq:ent_class:mixedness_transition_state} (i.e., we do not fix a random pure state $\ket{\psi}$, but generate new $\ket{\psi}$ for each $\lambda_i$). For each of these states, the random correlator dataset is calculated.
As before, we restrict to the second moment, $t=2$ [cf. Eq.~\eqref{eq:ent_class:statistical_moments}].
Since we want to visualize the continuous transition described by Eq.~\eqref{eq:ent_class:mixedness_transition_state}, we choose the UMAP parameters $n_\mathrm{neighbours} = 700$ and $d_\mathrm{min} = 0.8$.
This combination of a large number of nearest neighbours and a large minimal distance favours the formation of a single connected component in the embedding and prevents clustering, as discussed in Sec.~\ref{sec:ent_class:manifoldlearning} \footnote{For $n_\mathrm{neighbours}$ of the order $\sim 100$ and below, the embedded geometry in Fig.~\ref{fig:ent_class:mixednetss_transition} starts to fall apart into several unconnected components.}.
The results are shown in Fig.~\ref{fig:ent_class:mixednetss_transition} for $N=6$ qubits and two exemplary partitions $P$.
The embedded points are arranged  in serpentine shapes for both partitions.
In the first and third panel, the data points are colour coded according to the random state's purity $\gamma = \tr \rho_i^2$.
The purity continuously decreases along the serpentine shape and is thus accurately captured by the low-dimensional UMAP embedding.
In addition, the random correlator data point which represents the maximally mixed state $\id_d / d$ is embedded with the trained model, and indicated by a red cross. 
Note that, rather than by the convex combination  \eqref{eq:ent_class:mixedness_transition_state}, the transition of the state from pure to maximally mixed is also observed when progressive mixing is enforced by increasing $n_\mathrm{mixed}$ in \eqref{eq:ent_class:random_part_state}.

Since $\rho = \id_d / d$ is a fully separable state (with partition $[1,\dots, 1]$) located within the interior of the set of separable states \cite{zyczkowski_volume_1998,fine_equilibrium_2005,bengtsson_geometry_2006}, there is a finite distance from $\id_d / d$ within which all states are separable.
The UMAP embedding (which is a continuous function of the state) must exhibit this transition between entangled and separable states (more precisely PPT-states, as discussed below) along the serpentine shapes displayed in Fig.~\ref{fig:ent_class:mixednetss_transition}.

To quantify the entanglement of $\rho$ for a given partition $P$, we propose a novel, multipartite variant of the \textit{logarithmic negativity}. The logarithmic negativity, introduced in \cite{vidal_computable_2002}, is a mixed-state bipartite entanglement monotone \cite{plenio_logarithmic_2005} based on the positive-partial-transpose (PPT) criterion \cite{peres_separability_1996,horodecki_separability_1996}. It is given, for a bipartition $(A,B)$, by
\begin{equation}
	E_{\mathcal{N}, A} (\rho) = \log_2  \Vert \rho^{\Gamma_A} \Vert_1 \,.
\end{equation}
Here, $\Vert X \Vert_1 = \tr \sqrt{ X^\dagger X}$ is the trace norm, and $\rho^{\Gamma_A}$ describes the partial transpose of $\rho$ with respect to party $A$.
The logarithmic negativity $E_{\mathcal{N}, A} (\rho)$ is zero on the set of mixed separable states.
Positive $E_{\mathcal{N}, A} > 0$ implies entanglement between the parties $A$ and $B$. However, there also are entangled states with $E_{\mathcal{N}, A} = 0$. The set of states with $E_{\mathcal{N}, A} = 0$, commonly referred to as \textit{PPT states}, therefore includes the set of separable states \cite{bengtsson_geometry_2006}.
Based on the logarithmic negativity, we define the \textit{partition-log-negativity} \cite{brunner_interference_2023} for a given $P \in \mathbb{P}$ via
\begin{equation}\label{eq:ent_class:partition_log_negativity}
	E^P_\mathcal{N}(\rho) = \prod_{P_i \in P  \text{ if } |P_i| > 1} \Big( \tilde{E}_\mathcal{N} (\rho_{P_i})  \Big)^{\frac{|P_i|}{\Vert P\Vert} }\,,
\end{equation}
where $\rho_{P_i}$ is the reduced state of $\rho$ on the part $P_i$.
With $\Vert P\Vert$ we denote the number of entangled qubits in partition $P$ (i.e. not counting the qubits that are separable from all other qubits).
For example we have $\big\Vert [[0],[1,2,3,4,5]] \big\Vert = 5$ and $\big\Vert [[0,1],[2,3,4,5]] \big\Vert = 6$.
The auxiliary function $\tilde{E}_\mathcal{N}(\varrho)$ is the geometric mean of logarithmic negativities, taken over all bipartitions of the state $\varrho$,
\begin{equation}\label{eq:ent_class:part_log_negativity}
	\tilde E_\mathcal{N} (\varrho) = \left( \prod_{\text{bipartitions $(A,B)$ of $\varrho$}} E_{\mathcal{N}, A} (\varrho) \right)^{1/C} \,,
\end{equation}
where $C$ is the number of such bipartitions.
The partition-log-negativity constitutes a heuristic quantifier of entanglement within a given partition $P$.
It is given by the consecutive (weighted) geometric means over all parts of $P$ [Eq.~\eqref{eq:ent_class:partition_log_negativity}] and all bipartitions of each part [Eq.~\eqref{eq:ent_class:part_log_negativity}].
This definition ensures that a state $\rho$ with $E^P_{\mathcal{N}} (\rho) > 0$ exhibits non-vanishing logarithmic negativity for each bipartition of each part $P_i$ with $|P_i|>1$ of $P$. This implies that $\rho$ has entanglement partition $P$.
States with $E^P_\mathcal{N}(\rho) = 0$ (a proper superset of the separable states) have a positive partial transpose \cite{bengtsson_geometry_2006} for at least one bipartition of one of the parts $P_i$ and, hence, will again be called PPT-states (for this given partition). States which do not belong to this set are called NPT states for the given partition.

In the second and fourth panel of Fig.~\ref{fig:ent_class:mixednetss_transition}, we show the same embedded points as before, now coloured according to their partition-log-negativity $E^P_\mathcal{N}$.
PPT-states, with $E^P_\mathcal{N} (\rho) = 0$, are coloured black.
For both partitions, most states are entangled.
Moreover, we observe a clear transition between PPT states and NPT states.
Since the separable states are contained in the PPT states, the actual transition to the separable states must occur within the set of PPT states.
Our proposed entanglement quantifier $E^P_\mathcal{N} (\rho)$ is continuously and monotonically captured in the low-dimensional embedding.
This also demonstrates that, as expected, entanglement is stronger for increasing purity of $\rho$.
Most importantly, however, Fig.~\ref{fig:ent_class:mixednetss_transition} shows that there is a large neighbourhood of (mixed) entangled states (with $E^P_\mathcal{N} > 0$) around the pure entangled states ($\gamma \rightarrow 1$).
In this region, the mixedness of $\rho$ is insufficient to eliminate multipartite entanglement, such that the entanglement partition, and its detection via the UMAP embedding, is stable under noise.
Embedding new, unseen data points, for example stemming from real quantum computers, with the trained UMAP models, can be performed within a few seconds, providing an efficient scheme to characterise mixedness of multipartite entangled states.

Note that for partitions other than those shown in Fig~\ref{fig:ent_class:mixednetss_transition}, we obtain qualitatively similar results \cite{brunner_interference_2023,new_paper_inprep}. However, for partitions with a small entanglement depth (i.e. fewer entangled qubits), the transition between PPT states and NPT states is not as sharp.

\section{Conclusion}

In this contribution, we have demonstrated that the distributions of correlators measured in randomly chosen single-qubit bases contain sufficient information to assess the multipartite entanglement structure and the degree of mixedness of the (random) multipartite quantum states under scrutiny.
To extract distinctive features from this high-dimensional dataset, we employed a machine learning pipeline, based on the UMAP manifold learning embedding \cite{mcinnes_umap_2020}. 
Building on this statistical treatment, we showed that the second moment of the random correlators' distributions is sufficient to resolve the vast number of distinct entanglement partitions of an $N$-qubit system (Figs.~\ref{fig:ent_class:moment_comparison} and \ref{fig:ent_class:all_partitions}).
Moreover, correlators of order $N/2$ typically contain the largest amount of information to distinguish between the various partitions (see Fig.~\ref{fig:ent_class:correlation_order_comparison}).
Furthermore, we investigated the transition of entangled states of a given partition into the maximally mixed state, by sampling convex combinations of random pure states of the given partition with the maximally mixed state.
To quantitatively study the entanglement of states according to a given partition, we introduced a novel, heuristic mixed-state multipartite entanglement quantifier, the partition-log-negativity (based on the logarithmic negativity \cite{vidal_computable_2002}).
In the low-dimensional UMAP embedding, we identified a finite neighbourhood of the maximally mixed state outside of which all states are entangled (see Fig.~\ref{fig:ent_class:mixednetss_transition}), as expected from topological considerations \cite{zyczkowski_volume_1998,fine_equilibrium_2005,bengtsson_geometry_2006}.
Away from the associated separability threshold, the mixedness of the measured quantum states (induced by noise in the state generation circuit) does not suffice to erase multipartite entanglement.

In addition to the crossover between a given partition and the maximally mixed state, our method can also be generalised to transitions between distinct entanglement partitions by training the algorithm on convex interpolations, similar to Eq.~\eqref{eq:ent_class:mixedness_transition_state}, between random states with distinct partitions \cite{brunner_interference_2023}.
Furthermore, it would be interesting to complement our present statistical treatment by investigating the behaviour of structural entanglement properties, such as entanglement depth, e.g. monitored via the structural witnesses of \cite{lu_entanglement_2018}, along these crossovers between distinct partitions.

Regarding the scalability of our protocol, we found that the number of required measurements scales exponentially in the correlation order $k$. Since the best discrimination performance is shown by correlators of order $k = N/2$, our scheme is not applicable to analyse \textit{all} entanglement partitions for arbitrarily large systems, which, in any case, exhibit a super-exponentially large number of partitions [cf. discussion after Eq.~\eqref{eq:ent_class:random_pure_part_state}]. Note, however, that lower order correlators already achieve reasonably high accuracy scores in Fig.~\ref{fig:ent_class:correlation_order_comparison}.
Furthermore, we can train our model on data stemming from a system with $N_0 < N$ qubits and deploy it to predict entanglement partitions of all subsets of up to $N_0$ qubits of a large $N$-qubit system.
The generation of training data and the model training only depend on $N_0$, but not on $N$. Furthermore, the number of subsets of size $N_0$ scales polynomially in $N$, $\binom{N}{N_0} \sim N^{N_0}$.
In this way, we obtain a scalable scheme to characterise entanglement partitions in all subsets of up to $N_0$ out of $N$ qubits.



Our proposed method yields a versatile toolbox for the statistical analysis of mixed-state multipartite entanglement structures from data  readily accessible (requiring only single-qubits rotations and readout) in state-of-the-art quantum computing platforms.
The trained models provide an efficient characterisation tool to assess entanglement structure and mixedness of new, unseen data stemming from actual quantum hardware.
In \cite{new_paper_inprep}, we apply these models to characterise the multipartite entanglement properties of states prepared on actual quantum hardware, which are currently still strongly affected by noise. We find that we can correctly identify entanglement partitions of various states with high accuracy, and that the proposed scheme is able to detect entanglement even for strongly mixed states stemming from deep quantum circuits.
In general, the resolution of entanglement partitions (in the presence of noise) provides a detailed assessment of the multipartite entanglement structure of the computational state at the output of a quantum algorithm, allowing to identify qubits that have become entangled during the process.
Given the current generation of error-prone and noisy hardware, such characterisation schemes are crucial for the further development of the field.

\begin{acknowledgments}

A.X. acknowledges funding from the RISE Germany program of the German Academic Exchange Service (DAAD). The authors acknowledge support by the state of Baden-Württemberg through bwHPC and the German Research Foundation (DFG) through grants no INST 40/575-1 FUGG (JUSTUS 2 cluster) and no 402552777.

\end{acknowledgments}

\bibliography{entanglement_classification.bib}

\end{document}